\documentclass[a4,useAMS,usenatbib,usegraphicx]{mn2e}

\def\lsim{\mathrel{\rlap{\lower4pt\hbox{\hskip1pt$\sim$}}
    \raise1pt\hbox{$<$}}}                % less than or approx. symbol
\def\gsim{\mathrel{\rlap{\lower4pt\hbox{\hskip1pt$\sim$}}
    \raise1pt\hbox{$>$}}}                % greater than or approx. symbol

\def\apj{ApJ}
\def\apjl{ApJL}
\def\mnras{MNRAS}
\def\pasp{PASP}
\def\aj{AJ}

\def\aap{A\&A}
\def\apjs{ApJS}
\def\nat{Nature}

  \title[UV Dust Attenuation in Star-Forming Galaxies II]{UV Dust Attenuation in Star-forming Galaxies: II Calibrating the $A(UV)$ vs. $L_{TIR}/L_{UV}$ relation}

  \author[L. Cortese et al.]{L. Cortese$^1$, A. Boselli$^2$, P. Franzetti$^3$, R. Decarli$^{4,5}$, G. Gavazzi$^4$, S.Boissier$^2$, V. Buat$^2$\\
  $^1$ School of Physics and Astronomy, Cardiff University, Cardiff CF24 3AA, UK\\
  $^2$ Laboratoire d'Astrophysique de Marseille, Traverse du Siphon, BP8 13375 Marseille, France \\
  $^3$ INAF - IASF Milano, via Bassini 15, 20133 Milano, Italy\\
  $^4$ Universit\'{a} degli Studi di Milano-Bicocca, Piazza della Scienza 3, 20126 Milano, Italy\\ 
 $^5$ Universit\'{a} degli Studi dell'Insubria, via Valleggio 11, 22100 Como, Italy
    }

\begin{document}
\date{Accepted 2008 February 15.
      Received 2008 February 15;
      in original form 2007 October 12}
\pagerange{\pageref{firstpage}--\pageref{lastpage}} \pubyear{2002}

\maketitle

\label{firstpage}

\begin{abstract}
We investigate the dependence of the total-infrared (TIR) to UV luminosity ratio method for 
calculating the UV dust attenuation $A(UV)$ from the age of the underlying stellar populations by
using a library of spectral energy distributions for galaxies with different star formation
histories. Our analysis confirms that the $TIR/UV$ vs. $A(UV)$ relation varies significantly 
with the age of the underlying stellar population: i.e. for the same $TIR/UV$ ratio, systems with low specific star formation rate (SSFR) suffer a lower UV attenuation than starbursts. 
Using a sample of nearby field and cluster spiral galaxies we show that the use of a standard (i.e. age independent) $TIR/UV$ vs. $A(UV)$ relation leads to a systematic overestimate up to 2 magnitudes of the 
amount of UV dust attenuation suffered by objects with low SSFR and in particular HI-deficient 
star forming cluster galaxies.
This result points out that the age independent $TIR/UV$ vs. $A(UV)$ relation cannot be used to study the 
UV properties of large samples of galaxies including low star-forming systems and passive spirals.
Therefore we give some simple empirical relations from which the UV attenuation can be estimated taking into account 
its dependence on the age of the stellar populations, providing a less biased view of UV properties of  galaxies.

\end{abstract}


\section{Introduction}
The use of ultraviolet emission to shed light on the evolutionary history 
of galaxies is not straightforward.
The presence of dust in galaxies represents one
of the major obstacles complicating a direct quantification of the star formation activity
in local and high redshift galaxies.
Absorption by dust grains reddens the spectra at short wavelengths 
completely modifying the spectral energy distribution of galaxies.
Since the UV radiation is preferentially emitted by young stars 
that are generally more affected by attenuation from surrounding dust clouds 
than older stellar populations, rest-frame UV observations can lead to incomplete and/or biased reconstructions 
of the star formation activity and star formation history of 
galaxies affected by dust absorption, unless proper corrections are applied.\\
Radiative transfer models suggest that the total-IR (TIR) to UV luminosity ratio method (i.e. \citealp{buat92,xu95,meurer95,meurer99})
is the most reliable estimator of the dust attenuation in star-forming galaxies
because it is almost completely independent of the extinction mechanisms
(i.e. dust/star geometry, extinction law, see \citealp{buat96,meurer99,gordon00,witt00}).
This method is based  on the assumption that a fraction of photons emitted by newly formed young stars are 
absorbed by the dust. The dust heats up and subsequently re-emits the energy in the mid- and far-infrared.
The amount of UV attenuation can thus be quantified by means of an energy balance. 
The reliability of this method has also made the $TIR/UV$ ratio ideal 
to calibrate empirical methods to correct for dust attenuation when far-infrared observations 
are not available, like the well known $TIR/UV-\beta$ relation (\citealp{meurer99}, where $\beta$ is 
the slope of the UV continuum spectrum in the range 1300$<\lambda<$2600 \AA).\\
Although almost independent from dust-geometry and extinction law, the $TIR/UV$ ratio method unfortunately depends on the age of the underlying stellar populations.  
In systems with low specific star formation rate (SSFR) the dust heating by old stars becomes important, and only a fraction of the dust emission is related to the ultraviolet absorption \citep{gordon00,kong04,buat05}.\\
In the past this \emph{age effect} has been generally considered 
negligible since UV observations were available only for active star-forming systems for which 
the $TIR/UV$ vs. $A(UV)$ can be assumed to be independent from the star formation history (SFH, \citealp{gordon00,witt00}). However after the launch of the Galaxy Evolution Explorer (GALEX, \citealp{martin05}) this may not be the case any more. GALEX is delivering to the community an unprecedented amount 
of UV data on local and high redshift galaxies covering the whole range of morphologies and 
star formation histories: from starbursts, to passive spirals and elliptical galaxies (e.g. \citealp{n4438,bosell05,n4569,atlas2006,donas06}).

Can we extend the use of the \emph{standard method} (calibrated on active star forming galaxies and starbursts) to 
correct for dust attenuation to the thousands of galaxies detected by GALEX? 
In recent years various studies have been undertaken to address these issues (i.e. \citealp{kong04,denis05a,COdust05,atlas2006,panuzzo07,dale07}). 
However more attention has been always given to the $TIR/UV$ vs. $\beta$ relation than 
to the $TIR/UV$ vs. $A(UV)$ relation. 
\cite{kong04} suggested that quiescent galaxies deviate from the $TIR/UV-\beta$ relation of starburst galaxies, because they tend to have redder ultraviolet spectra at fixed $TIR/UV$ ratio. They interpreted the different behavior of starbursts and normal galaxies as due to a difference in the star formation histories, proposing that the offset from the starburst $TIR/UV-\beta$ relation can be predicted using the birthrate parameter $b$ (\citealp{kennbirth}; i.e. the ratio of the current to the mean past star formation activity). 

In this work we use a different approach, investigating the dependence of the 
$TIR/UV$ vs. $A(UV)$ relation on the galaxy star formation history, to quantify the 
influence of the age-independent correction on the interpretation of 
UV observations.
We will adopt a very simple dust geometry model and extinction law to show that the use 
of standard (i.e. age independent) methods can strongly bias our interpretation of UV 
properties of local and high 
redshift galaxies. 
The main goal of this paper is to provide the community with some new empirical 
relations based  on observable quantities suitable for deriving dust attenuation corrections  
taking into account the \emph{age effect}.
More detailed geometries and dust models have been developed in the last decade (e.g. \citealp{buat96,silva98,bianchi00,charlfall00,witt00,calzetti01,panuzzo03,piovan06}). 
However, since our results do not depend on the geometry or extinction law adopted (as shown in Appendix A), we decided 
to adopt a very simple approach in order to make our recipes useful for the widest possible range of 
astrophysical applications.

In Section 2 we present our model and discuss the variation of the $TIR/FUV$ vs. $A(FUV)$ relation with 
the age of the underlying stellar population. In Section 3 we apply our model to a sample of nearby field and cluster 
galaxies and in Section 4 we use this sample to test new empirical recipes to determine $A(FUV)$. Additional 
applications of these techniques are discussed in Section 5.

\section{The model}
\label{model}

In order to investigate the correlation between the UV dust attenuation $A(UV)$ and the $TIR/UV$ ratio 
for different stellar populations we used a library of spectral energy distributions (SEDs) obtained 
using the Bruzual \& Charlot population synthesis models \citep{BC2003}.
We adopted a \cite{salpeter} initial mass function (IMF) and a star formation history (SFH) 'a la Sandage' in the formalism used by \cite{gav02}:
\begin{equation} 
SFR(t,\tau)=\frac{t}{\tau^{2}}\times\exp(-\frac{t^{2}}{2\tau^{2}})  
\end{equation}
where $SFR$ is the star formation rate per unit mass, $t$ is the age of the galaxy (we assumed $t$=13 Gyr at the present epoch) and $\tau$ is the time at which the star formation rate reaches the highest value over the whole galaxy history: 
short $\tau$ correspond to galaxies dominated by old stellar populations while long $\tau$ correspond to young (i.e. star forming) galaxies. 
We investigated a range of $\tau$
between 0.1 and 25 Gyr, with steps of 0.2 Gyr  for 0.1$<\tau<$10 Gyr and 0.5 Gyr for $\tau>$10 Gyr and metallicities in the range 0.02$\leq Z \leq$2.5 Z$_{\odot}$ 
in five steps: 0.02, 0.2, 0.4, 1, and 2.5 Z$_{\odot}$.
The library obtained is able to reproduce all the SEDs typically observed in local galaxies \citep{gav02} 
and it is only a function of $\tau$\footnote{Throughout the paper, $\tau$ will be only used to indicate 
the shape of the SED and it must not be adopted to quantify the real age of the stellar populations.}. Similar libraries can be obtained assuming different SFH (e.g. 
exponential SFH, see Appendix A).\\
Each synthetic SED in our library has been reddened using different values 
of $A(FUV)$ in the range 0$<A(FUV)<$ 4 mag, the typical range observed in normal star-forming 
galaxies (e.g. \citealp{buat02,atlas2006,boissier06}). Higher UV attenuations are normally associated with strong starbursts and highly obscured 
objects which are outside the goal of the present work. 
For each $A(FUV)$, $A(\lambda)$ 
(the attenuation at each $\lambda$) has been derived assuming a Large Magellanic Cloud (LMC) extinction curve \citep{pei92}:
\begin{equation}
\small
k(\lambda) =\left\{\begin{array}{ll}
1.962\times(\lambda^{-1})-0.55 & \textrm{if $\lambda^{-1}<$4.2 $\rm \mu m^{-1}$}\\
~~\\
-375.91+231.23\times(\lambda^{-1})- & \textrm{if ($\lambda^{-1}>$4.2 $\rm \mu m^{-1}$ \&}\\
-46.204\times(\lambda^{-2}) +3.0721\times(\lambda^{-3}) &  \textrm{ $\lambda^{-1}<$5.5 $\rm \mu m^{-1}$)}\\
~~\\
1.694\times(\lambda^{-1})-0.20 & \textrm{if $\lambda^{-1}>$5.5 $\rm \mu m^{-1}$}\\
\end{array}\right.
\end{equation}
and a simple sandwich model for dust geometry \citep{bosellised}, where a thin layer of dust of thickness $\zeta$ 
is embedded in a thick layer of stars:
\begin{eqnarray}
\label{alambda}
A(\lambda)=-2.5 \cdot \log\bigg(\left[\frac{1-\zeta(\lambda)}{2}\right]\left(1+{\rm e}^{-\tau_{dust}(\lambda) \cdot {\rm sec}(i)}\right) + &&{} \nonumber \\
{} +\left[\frac{\zeta(\lambda)}{\tau_{dust}(\lambda) \cdot {\rm sec}(i)}\right] \cdot \left(1-{\rm e}^{-\tau_{dust}(\lambda) \cdot {\rm sec}(i)}\right) \bigg)   ~~[{\rm mag}]
\end{eqnarray}
where $i$ is the galaxy inclination, $\tau_{dust}(\lambda)$ is the optical depth and the dust to stars scale height ratio $\zeta(\lambda)$ depends 
on $\lambda$ (in units of \AA) as \citep{bosellised}:
\begin{equation}
\label{zeta}
\zeta(\lambda)=1.0867{-}5.501 \times 10^{-5} \times \lambda
\end{equation}
In the case of the FUV band ($\lambda\sim1530$ \AA) $\zeta\sim1$. 
In this case $\tau_{dust}$(FUV) can be derived by inverting Eq. \ref{alambda}:
\begin{eqnarray}
\tau_{dust}({\rm FUV})=[1/{\rm sec}(i)] \cdot \big(0.0259+1.2002 \times A({\rm FUV})  + 1.5543 \times  &&{} \nonumber \\	 
 {}  \times A({\rm FUV})^2-0.7409 \times A({\rm FUV})^3 +0.2246 \times A({\rm FUV})^4 \big) &&
\end{eqnarray}
Using the LMC extinction law $k(\lambda)$, we then derive:
\begin{equation}
\tau_{dust}(\lambda) = \tau_{dust}({\rm FUV}) \cdot k(\lambda) / k({\rm FUV}) 
\end{equation}
and compute the complete set of $A(\lambda)$ using Eq.\ref{alambda}.
In order to simplify the calculations in the following we assume $i$=0.
We adopted the LMC extinction law as a compromise between a Milky Way 
extinction law, with a strong bump at 2175 \AA, and the Small Magellanic Cloud, with almost 
no bump at $\sim$2000 \AA. Recent analysis have also pointed out 
that a LMC extinction law can reproduce GALEX observations better than a MW or SMC extinction 
law \citep{denis05a,inoue06}

\begin{figure}
\centering
\includegraphics[width=8.5cm]{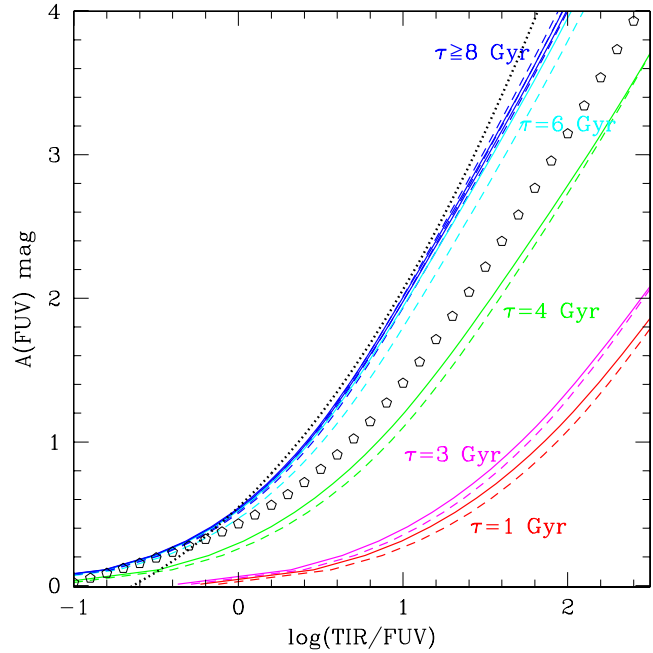}
\caption{\label{tauafuv}The relationship between the $TIR/FUV$ ratio and the 
FUV attenuation $A(FUV)$ obtained from our model for different values of $\tau$. 
Solid and dashed lines show the relations for stellar metallicity $Z$=2.5 Z$_{\odot}$ and $Z$=0.02 Z$_{\odot}$ respectively. 
The dotted black line indicates the age independent relation given by Buat et al. (2005). 
Empty pentagons show the relation proposed by Kong et al. (2004) for birthrate parameter 
$b\sim$0.06.}
\end{figure}

The reddened SED is then subtracted from 
the original \emph{dust-free} SED, providing (once integrated over all wavelengths) the amount of energy absorbed by dust
which, assuming an energetic balance, corresponds to the total infrared radiation (TIR) emitted by the galaxy. Finally, by convolving the reddened SED with the GALEX-FUV filter we estimated the FUV flux and the $TIR/FUV$ ratio for each model.
The relations between $A(FUV)$ and $TIR/FUV$ so obtained for different values of $\tau$ and stellar metallicity are shown in Fig.\ref{tauafuv}.\\
The mean age of the stellar populations plays a crucial role in the $TIR/FUV$ vs $A(FUV)$  relation, as already pointed out by several theoretical (e.g. \citealp{buat96,vero98,witt00,gordon00,kong04,buat05}) and 
observational (e.g. \citealp{sauvage92,walterbos,popescu02}) studies: for the same $TIR/FUV$ ratio
active star forming galaxies ($\tau\geq$7 Gyr) are more attenuated than 
objects with low SSFR ($\tau\sim$ 4-5 Gyr) or quiescent galaxies ($\tau \sim$ 2 Gyr).
On comparison, we find a very weak dependence ($<$0.1 mag) of the $TIR/FUV$ vs. $A(FUV)$ on stellar metallicity (see Fig.\ref{tauafuv}). 
In Table \ref{convFUV} we provide the best polynomial fit (averaged over the whole metallicity range considered) to 
the $A(FUV)$ and $TIR/FUV$ for different values of $\tau$.

\begin{figure}
\centering
\includegraphics[width=8.5cm]{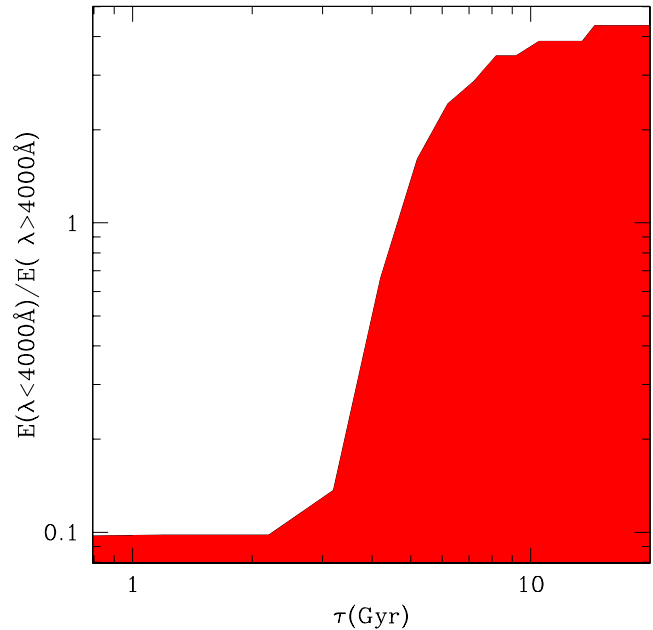}
\caption{\label{enbudg}The importance of the UV emission in the dust heating. The ratio of the 
energy absorbed by dust at $\lambda<4000 \rm\AA$ and at $\lambda>4000\rm \AA$ is shown as 
a function of $\tau$ (i.e. the age of the Universe at which the galaxy SFR peaks).}
\end{figure}

The origin of the \emph{age effect} is clearly visible in Fig.\ref{enbudg} 
where we compare the amount of the energy absorbed by dust 
at long ($\lambda>4000 \rm \AA$) and short ($\lambda<4000 \rm \AA$) wavelengths for different 
values of $\tau$. 
For $\tau\leq 5$ Gyr the high-energy photons (mainly UV) contribute to less than the 50\% of 
the total energy absorbed by dust and the far infrared emission is mainly due to the re-emission of the 
stellar radiation emitted by intermediate age stars in the optical. 
Only for $\tau\geq6-7$ Gyr the UV radiation dominates the dust heating, contributing more than the 75\% 
of the whole energy absorbed and re-emitted in the far infrared.
This also implies that the far-infrared radiation is not always a good star formation indicator. \\ 
We remark that our results do not strongly depend on the parameters adopted in our model (e.g. dust geometry, attenuation law). The variation in the $TIR/FUV$ vs. $A(FUV)$ relations when different attenuation laws are considered is $\leq$0.2 mag, as discussed in Appendix A.
This is confirmed by the good agreement between our model and the one 
of \cite{buat05} (black dotted line in Fig.\ref{tauafuv}) in the case of strong star forming systems 
and starburst galaxies ($\tau\geq$8 Gyr).  
Our results are also independent of the shape of the adopted SFH and galaxy age. 
In fact the $TIR/FUV$ vs. $A(FUV)$ relation mainly depends on the specific star-formation and, 
for the same specific star formation activity, it is not strongly affected by the real 
shape of the past SFH (see Appendix A).
Moreover, in Fig.\ref{tauafuv} we show the $TIR/FUV$ vs. $A(FUV)$ relation proposed by \cite{kong04} (empty pentagons) for a birthrate parameter $b\sim$0.06 (the lowest birthrate parameter considered in their model, corresponding to $\tau\sim$4.1 Gyr in our formalism). The level of agreement ($\sim$0.2 mag) between \cite{kong04} and our model is quite comforting, considering that the \cite{kong04} model is based on different SFH (i.e. exponential+bursty model) and geometry (i.e. the time-dependent scenario proposed by \cite{charlfall00}).\\
As already discussed by \cite{gordon00}, the flux ratio method would not be applicable in the case of an embedded starburst in 
a galaxy with a second older, less embedded stellar population. 
At UV and IR wavelengths, the starburst would dominate, but at optical and near-IR wavelengths the older population would dominate.
However, at least for spiral galaxies in the local Universe this seems not to be the case \citep{jorgehauv} and smooth SFH like an 'a la Sandage' 
and an exponential well reproduce the observations (e.g. \citealp{boselli,gav02,heavens,panter07}).
We therefore conclude that our results are model-independent within 0.2 mag (see Appendix A).

\section{Application to normal star-forming galaxies.}
\label{data}
In order to quantify the impact of the \emph{age effect} on the estimate of the UV attenuation,  
we computed $A(FUV)$ for a sample of spiral galaxies and compared our estimate with the 
one obtained using the standard relation calibrated on active star-forming galaxies (e.g. $\tau\geq$8 Gyr, $FUV-H\leq$4 mag, $b\geq$0.5).
  
\subsection{The data}
The sample here adopted is 
an extension of the optically selected sample described in \cite{COdust05}, 
composed by late-type galaxies (later than S0a) including 
giant and dwarf systems extracted from the Virgo Cluster Catalogue (VCC, \citealp{vcc}) 
and from the CGCG catalogue \citep{ZWHE61}.
The data include galaxies in the Virgo, Abell1367 and Abell262 clusters and part of the Coma-A1367 supercluster (including cluster and field galaxies) observed as part 
of the All-sky Imaging Survey (AIS) and of the Nearby Galaxy Survey (NGS) carried out by GALEX 
in two UV bands: FUV ($\rm \lambda_{eff}=1530\AA, \Delta \lambda=400\AA$) 
and NUV ($\rm \lambda_{eff}=2310\AA, \Delta \lambda=1000\AA$).
We include in our analysis only late-type galaxies 
detected in both NUV and FUV GALEX bands and in 
both 60 $\mu m$ and 100 $\mu m$ IRAS bands: 191 galaxies in total ($\sim$70\% in 
high density environments).
UV and far-infrared data have been combined with multifrequency data available.
These are optical and near-IR H imaging \citep{gav00,bosellised}, 
most of which are available from the 
GOLDMine galaxy database \citep{goldmine} (http:\slash \slash goldmine.mib.infn.it).
Data from UV to near-IR have been corrected for Galactic extinction
according to \cite{burnstein82}.\\
We assume a distance of 17 Mpc for the members
of Virgo Cluster A, 22 Mpc for Virgo Cluster B, and 32 Mpc for objects
in the M and W clouds \citep{gav99}.
Members of the Cancer, A1367, and Coma clusters are assumed to
lie at distances of 65.2, 91.3, and 96 Mpc, respectively.
Isolated galaxies in the Coma supercluster are assumed 
at their redshift distance, adopting $H_{0}$ = 75 $\rm km~s^{-1}~Mpc^{-1}$.

\subsection{SED fitting technique}
The FUV attenuation of each galaxy in our sample has been computed 
using the following SED fitting procedure,  implemented into the Galaxy Observed Simulated SED Interactive Program 
(GOSSIP, \citealp{paolotesi,franzetti07}).
In order to take into account the \emph{age effect} we reddened each synthetic SED and fit them to 
the observed FUV-to-near-infrared SED obtained using all
the available photometric magnitudes.\\
In details, for each galaxy we compute the observed $TIR/FUV$ ratio using IRAS and GALEX 
observations.
The TIR flux emitted in the
range 1-1000 $\mu m$, is obtained following \cite{dale01}:
\begin{eqnarray}
 \label{tir}
 \log(f_{TIR})  =  \log(f_{FIR}) + 0.2738 - 0.0282\times\log(\frac{f_{60}}{f_{100}})+
 \nonumber\\
 +0.7281\times\log(\frac{f_{60}}{f_{100}})^{2}+0.6208\times\log(\frac{f_{60}}{f_{100}})^{3} +
 \nonumber\\
 + 0.9118\times\log(\frac{f_{60}}{f_{100}})^{4}
 \end{eqnarray}
where $f_{FIR}$ is the far-infrared flux, defined as the flux between 42 and 122 $\mu m$ \citep{helou88}:
\begin{equation}
 f_{FIR} = 1.26 \times (2.58 \times f_{60} + f_{100}) \times 10^{-14} ~~[\rm Wm^{-2}] 
 \end{equation}
and $f_{60}$ and $f_{100}$ are the IRAS flux densities measured at 60 and 100 $\mu m$ (in Jansky).
Using the relations presented in the previous section, we then convert the observed $TIR/FUV$ 
ratio into $A(FUV)$ for each value of $\tau$ and $Z$ considered in our model and determine $A(\lambda)$ as described in Sec.\ref{model}.
Each \cite{BC2003} SED is then reddened with the $A(\lambda)$ obtained following this procedure and 
fitted to the observed (i.e. not corrected for internal extinction) SED, using 
a $\chi^{2}$ fitting technique similar to the one described in \cite{gav02} and 
assuming a conservative photometric error of 0.15 mag for each band.

For each galaxy, the value of $\chi^{2}$ determines the weight of a given model (as $\exp(-\chi^{2}/2)$) and a probability 
distribution function (PDF) for $\tau$ can be build by combining weights for each model. 
We then normalize the final PDFs and use the peak value (i.e. the most probable one) as 
the best estimate of $\tau$ and the range of $\tau$ containing 68.2\% of the PDF's area 
(equivalent to the use of constant $\chi^{2}$ contours) as estimate of the 1 $\sigma$ error. 
The SED fitting provides not only the best value of $\tau$ but also 
the right estimate of $A(FUV)$ and its uncertainty.
The combined use of the observed $TIR/FUV$ ratio for each galaxy and of the 
relations presented in the previous Section makes this method ideal to 
estimate the UV attenuation for a wide range of morphological types and star formation histories.

\subsection{The impact of the SFH on the estimate of $A(FUV)$. }
\begin{figure}
\centering
\includegraphics[width=8.5cm]{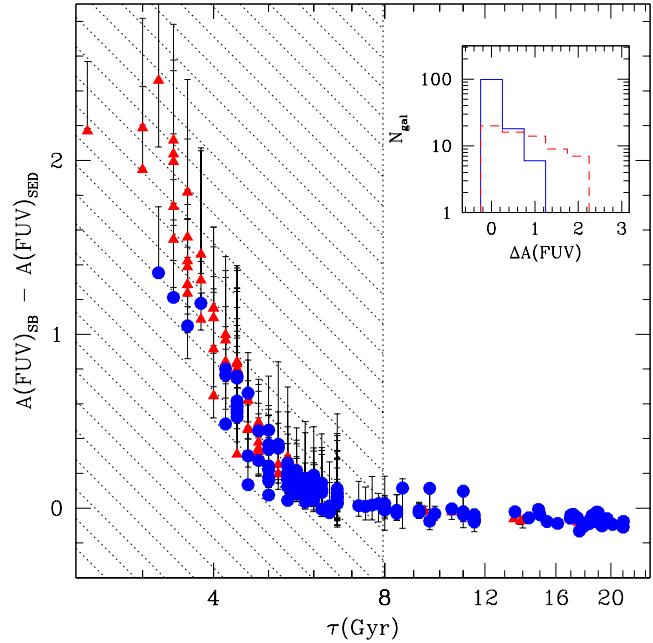}
\caption{\label{diffAFUV_tau} The difference between the FUV attenuation for active star forming galaxies ($\tau\geq$ 8 Gyr; consistent with Buat et al. (2005)) and the one obtained from our SED fitting technique as a function of $\tau$. Circles indicate healthy spirals (HI-DEF$<$0.4) and triangles HI-deficient star forming objects (HI-DEF$\geq$0.4). The distributions of the difference for healthy (solid histogram) and deficient (dashed histogram) spirals are shown in the upper-right panel. 
The unshaded area indicates the range of $\tau$ on which the standard $TIR/FUV$ vs. $A(FUV)$ relation for star-forming galaxies is usually calibrated.}
\end{figure}

\begin{figure}
\centering
\includegraphics[width=8cm]{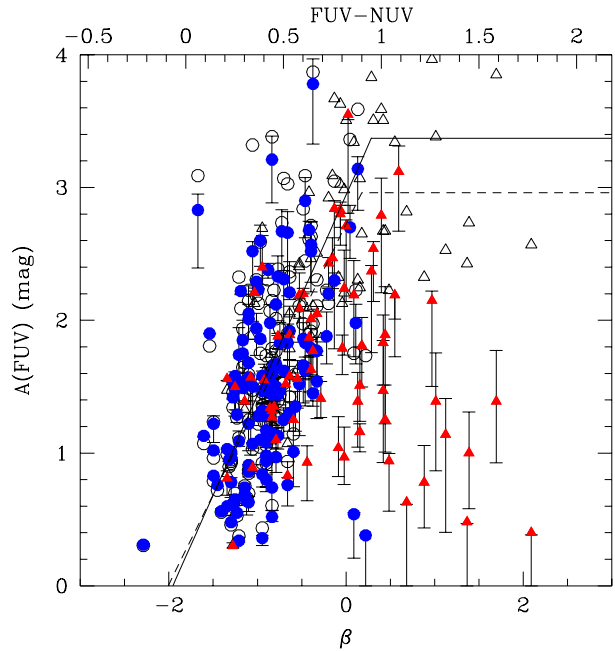}
\caption{\label{AFUV_beta} The $A(FUV)$ vs. $\beta$ relation for our optically selected sample.
A(FUV) is computed from our SED fitting technique (filled symbols) and using the standard recipe for active star forming galaxies (empty symbols) respectively. The solid and dashed lines indicate the relations between $A(FUV)$ and $FUV-NUV$ proposed by Salim et al. (2007). Symbols 
are as in Fig.\ref{diffAFUV_tau}.}
\end{figure}

The difference between the FUV attenuation obtained with the SED fitting technique ($A(FUV)_{SED}$) and the one 
obtained by blindly applying a constant conversion calibrated on star forming galaxies and starbursts ($\tau\geq$8 Gyr, $A(FUV)_{SB}$ consistent 
with the one presented by \citealp{buat05}) is show in Fig.\ref{diffAFUV_tau} as a function of $\tau$ (for clarity only 
the errors on A(FUV) are shown).
As already shown in Fig.\ref{tauafuv}, for low values of $\tau$ the standard conversion overestimate the UV dust attenuation: 
in particular for $\sim$ 30\% (59/191 galaxies) of our sample the use of a constant $TIR/FUV$ vs. $A(FUV)$ relation 
leads to an overestimate of more than 0.5 mag of the UV attenuation and for $\sim$16\% this 
systematic error exceeds 1 mag. 
In order to investigate the properties of the most deviating objects in Fig.\ref{diffAFUV_tau} we divide 
our sample according to their HI-deficiency\footnote{The 
HI deficiency is defined as the difference, in logarithmic units, between the 
observed HI mass and the value expected from an isolated galaxy 
with the same morphological type $T$ and optical linear diameter $D$:
HI-DEF = $<\log M_{HI}(T^{obs},D^{obs}_{opt})> - log M^{obs}_{HI}$ \citep{haynes}}: a value of $HI-DEF$ = 0.4 has been used to 
separate healthy (blue circles) from HI-deficient star forming spirals (red triangles).
HI-deficient spirals are among the most deviating objects in Fig.\ref{diffAFUV_tau}, consistent 
with the fact that gas deficient objects have normally a SSFR significantly lower than that expected from their luminosity \citep{boselli}. 
However also some healthy spirals show a significant offset from the \emph{age independent} $TIR/FUV$ vs. A(FUV) relation. These are massive early type spirals with low SSFR, similar to M31 in the Local Group (i.e. $FUV-H\sim$7 mag,  \citealp{atlas2006}). \\
The $TIR/FUV$ ratio cannot thus be considered as a good proxy of the UV attenuation for sample of galaxies 
spanning a wide range of SSFR, in particular in clusters, and relations like the $TIR/UV-\beta$ relation \citep{meurer99,kong04} cannot be blindly 
used to determine $A(FUV)$. This is shown in Fig. \ref{AFUV_beta} 
where we compare the $A(FUV)$ vs. $\beta$ relation obtained from our SED fitting technique (filled symbols) with the one obtained 
when $A(FUV)$ is computed using the $A(FUV)$ vs. $TIR/FUV$ relation for strong star forming galaxies  (empty symbols). 
The $\beta$ parameter has been computed from the FUV-NUV colour following \citep{kong04}: $\beta$=2.201$\times(FUV-NUV)$-1.804. 
Galaxies with $\beta>-0.2$ ($FUV-NUV>$0.75) are mainly HI-deficient galaxies and have a FUV dust attenuation $\sim$2 mag lower than the one obtained using the standard $A(FUV)-\beta$ relation: a difference significantly larger than the typical error on the estimate of A(FUV) ($\leq$0.5 mag) by our fitting technique.
The systematic errors introduced in the data is similar when the $TIR/FUV$ ratio is not available and $A(FUV)$ is determined using empirical methods based on the $FUV-NUV$ colour like the one proposed by \cite{salim07}, and once again calibrated 
on strong star forming galaxies (solid and dashed line in Fig.\ref{AFUV_beta}): 
for $\beta>$-0.2, a UV attenuation in the range  2.99$<A(FUV)<$3.32 mag is predicted, whereas our method gives an 
average value of $A(FUV)\sim$ 1 mag.
This is mainly due to the fact that in low star forming systems the UV spectral slope 
is strongly contaminated by the old stellar populations, whose contribution increases the value of $\beta$ even if the FUV attenuation is low.
This result is consistent with the recent analysis of a sample of $\sim$1000 galaxies selected from SDSS presented by \cite{johnson1,johnson2}, who found that a large fraction of galaxies having red $FUV-NUV$ colours is also characterized by large 4000 \AA ~break ($D(4000)$), suggesting that 
part of the dust heating comes from old stellar populations and not from extremely obscured star forming regions. 
We can therefore conclude that the $TIR/UV-\beta$ relation can be blindly used to estimate $A(FUV)$ only for $\beta<-0.2$ ($FUV-NUV<$0.75).

\section{Optimized $A(FUV)$ determination: recommended methods.}
\label{recipe}
It clearly appears that a correct estimate of the UV attenuation requires 
information about the shape of the galaxy SED. 
A SED fitting technique like the one here described represents therefore
the best method available to quantify $A(FUV)$ and properly correct UV observations.
Unfortunately SED fitting is only possible when large multiwavelength data sets are available, which is not always the case.
For this reason we investigated the possibility to derive different techniques which can be used 
to estimate $A(FUV)$ when SED fitting is not possible. 
Ideally, these techniques will have a small rms and not display the systematic shift relative 
to the galaxy age observed in Fig.\ref{diffAFUV_tau}.

\begin{figure}
\centering
\includegraphics[width=8cm]{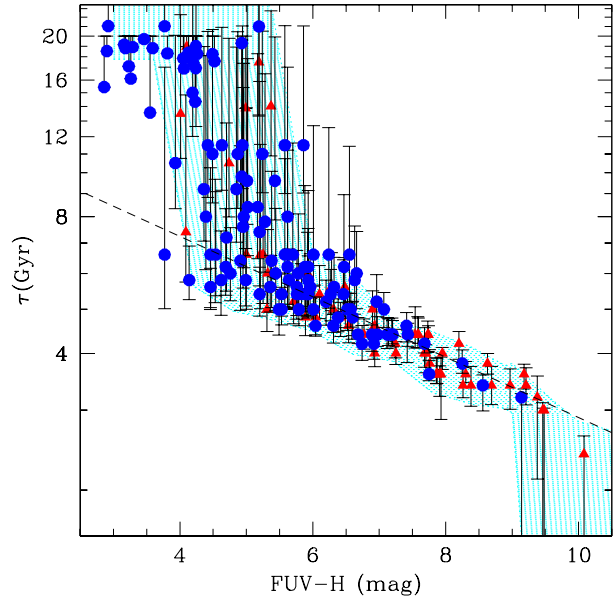}
\caption{\label{FUVH_tau} Relation between the observed (i.e. not corrected for internal extinction) $FUV-H$ colour and $\tau$ for 
our sample. The dotted line indicates the best linear fit for galaxies having $\tau\leq$7 Gyr. The shaded area indicate the typical 
dispersion of the data computed in FUV-H bins 0.5 mag wide. Symbols 
are as in Fig.\ref{diffAFUV_tau}.}
\end{figure}

\begin{figure}
\centering
\includegraphics[width=8cm]{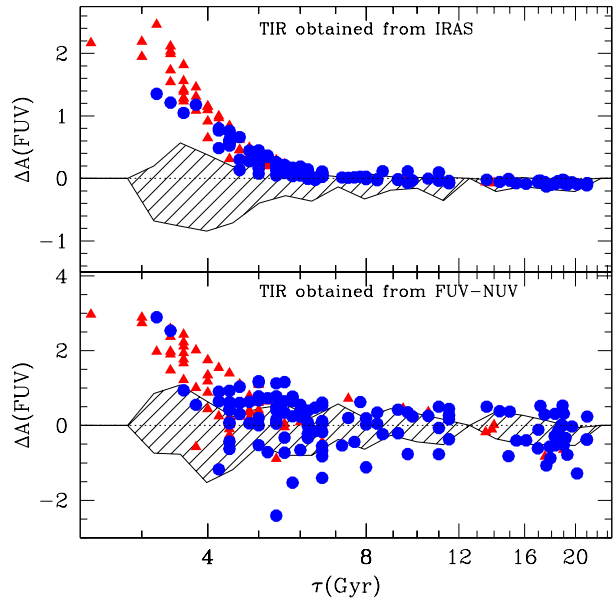}
\caption{\label{test} Residuals in the estimate of $A(FUV)$ obtained with the different methods described in Sec. 4.
The $TIR/FUV$ ratio is estimated using IRAS observations (upper panel) and the $FUV-NUV$ colour 
(lower panel) respectively. 
The shaded areas show the typical uncertainties in the new recipes presented in this paper, estimated by 
generating 1000 random galaxies having $FUV-H$ colour following a gaussian distribution 
centered on the observed colour with $\sigma$=0.2 mag (see Sec. 4 for details).
For comparison, the residuals obtained when an age independent method is used are indicated by the filled symbols. Symbols 
are as in Fig.\ref{diffAFUV_tau}.}
\end{figure}

The crucial step is to find a good proxy for $\tau$ (and therefore the shape of the SED).
This is particularly important for $\tau<6-7$ Gyr where the $TIR/FUV$ vs. $A(FUV)$ relation 
strongly depends on the age of the stellar populations. 
The first natural choice is to look for a colour with the largest possible dynamical range, sensitive to small changes in the shape of the SED \citep{gav02}.
Therefore in Fig.\ref{FUVH_tau} we plot the distribution of $\tau$ for our sample as a function of the observed (i.e. not corrected 
for internal extinction) $FUV-H$ colour.
For $\tau>$7 Gyr the two variables are not correlated and galaxies show approximately the same 
$FUV-H$ colour independently from the value of $\tau$, reflecting the large error on the estimate of $\tau$. 
This is as expected since in young stellar populations ($\tau>$7 Gyr) the 
variations in the observed $FUV-H$ are mainly due to dust attenuation and not to age.  
However for $\tau\leq$7 Gyr (i.e. the range in which we are interested) 
the $FUV-H$ is a very good proxy for $\tau$ (Pearson correlation coefficient $r\sim -0.91$). A simple lest-square fitting gives us:
\begin{equation}
\log(\tau) = -0.068 \times (FUV-H) + 1.13 
\end{equation}

which can be used to estimate $\tau$ from the $FUV-H$ colour.
The dispersion on this relation varies quite remarkably with the colour. We therefore determine the 
typical dispersion within 0.5 mag wide FUV-H bins by combining the $\tau$'s PDF for each galaxy in the bin and estimating the 
1$\sigma$ error as described in the previous section. The dispersion in the relation (indicated by the shaded area in Fig.\ref{FUVH_tau}) 
is $\Delta(log(\tau))\sim 0.12-0.15$ for 6$<FUV-H<$9 mag, increasing significantly (up to $\Delta(log(\tau))\sim$1) for redder or bluer colours.
Similar relations are found when we consider different far-ultraviolet-optical colours and are presented in Table \ref{taucol}.
We therefore propose two different ways to determine $A(FUV)$ depending on whether or not 
far infrared observations are available\footnote{A detailed guide to the recipes presented in this paper 
can also be found at http://www.astro.cf.ac.uk/pub/Luca.Cortese/UVattenuation.html.}.
In Appendix B we also provide similar recipes in order to determine $A(NUV)$ in the case FUV observations are not available.\\

i) {\bf Far-infrared observations are available.} Determine the observed $TIR/FUV$ ratio and use one of the relations 
presented in Table \ref{taucol} to determine $\tau$. Finally use the value of $\tau$ so obtained to choose the conversion between the $TIR/FUV$ ratio and $A(FUV)$ from the relations provided in Table \ref{convFUV}.
As discussed above we suggest using the colour covering the widest possible dynamical range, i.e. 
first choice $FUV-H$, last choice $FUV-g$.
The error on the estimate of the FUV dust attenuation ($\sigma(A(FUV))$) obtained with this method depends on the observational errors on the FUV-optical/near-infrared colour and on the intrinsic dispersion of the colour-$\tau$ relation adopted.
In order to estimate $\sigma(A(FUV))$, for each galaxy in our sample 
we generated 1000 random galaxies having $FUV-H$ colour following a gaussian distribution 
centered on the observed colour with $\sigma$=0.2 mag (more accurate estimate of the $FUV-H$ colour would correspond to a lower uncertainty 
on $A(FUV)$). 
For each $FUV-H$ colour the $\tau$'s PDF are used to random generate the correspondent value of $\tau$\footnote{This step is included in order 
to take into account the intrinsic dispersion of the $FUV-H$ vs $\tau$ relation in the estimate of $\tau$.} and then 
the final value of $A(FUV)$ by applying the relations in Table \ref{taucol}.
In Fig.\ref{test} (upper panel) we compare this method with the SED fitting technique.  
The shaded area indicates the 1 $\sigma$ error on the estimate of $A(FUV)$.
As expected, the error on the estimate of $A(FUV)$ varies significantly with $\tau$: from 
$\sim$+0.1/-0.25 mag for $\tau\geq$6 Gyr ($FUV-H<$5 mag) to a maximum of +0.4/-0.8 mag for $3.5\leq\tau\leq4.5$ Gyr (7$<FUV-H<$8.5 mag). 
Even if the uncertainty for low $\tau$ is quite large (due to the large variation of the $TIR/FUV$ vs. $A(FUV)$ relation with $\tau$), 
it is clear that our empirical method is able to remove the systematic overestimate of $A(FUV)$ 
when the age independent $TIR/FUV$ vs. $A(FUV)$ conversion is used (filled symbols in Fig.\ref{test}).\\

ii) {\bf Far-infrared observations are not available.} In this case the first step is 
to find a way to estimate the $TIR/FUV$ ratio from the available observations. Recently 
various methods have been proposed based on the use of the $FUV-NUV$ colour (or $\beta$, \citealp{seibert05,COdust05,boissier06,atlas2006}), H-band luminosity \citep{COdust05}, gas metallicity \citep{COdust05,boissier06}, effective surface brightness \citep{COdust05,johnson1}, 
$D(4000)$ or optical and ultraviolet colours \citep{denis05a,johnson06,johnson1}. 
Once the $TIR/FUV$ ratio has been determined it is possible to proceed as described above: i.e. estimate $\tau$ from an ultraviolet-optical/near infrared colour (see Table \ref{taucol}) and than convert $TIR/FUV$ in $A(FUV)$ using the proper 
relation in Table \ref{convFUV}.

To test this second method we estimate
the $TIR/FUV$ ratio from the $FUV-NUV$ colour \citep{COdust05}:
\begin{equation}
\small
\log(\frac{TIR}{FUV}) =\left\{\begin{array}{ll}
0.7 \times (2.201 \times FUV-NUV -1.804) + 1.3 \\
\textrm{if ($FUV-NUV\leq$0.9)}\\
~\\
1.424  \\
 \textrm{if ($FUV-NUV>$0.9)}\\
\end{array}\right.
\end{equation}
and then determine $\tau$ from the $FUV-H$ colour.
The typical uncertainty in this method has been computed following the same procedure described in the previous point.
Also in this case our recipe is able to remove the systematic overestimate 
on $A(FUV)$. This is clearly evident in Fig.\ref{test} (lower panel) where our method is compared with the age independent method based on the $FUV-NUV$ colour proposed by \cite{salim07}. 
Unfortunately the uncertainty on the determination of $A(FUV)$ considerably increases to 
$\sim$+0.5/-0.7 mag for $\tau\geq$6 Gyr ($FUV-H<$5 mag) reaching a maximum of $\sim$+1/-1.2 mag for $3.5\leq\tau\leq4.5$ Gyr (7$<FUV-H<$8.5 mag) when the total infrared luminosity is estimated from empirical relations involving colours or luminosities.
As discussed by \cite{COdust05} this is mainly due to the intrinsic scatter in 
the empirical relation used to determine the $TIR/FUV$ ratio.\\

\noindent
In the very unlikely scenario in which there is no sufficient data to estimate 
the $TIR/FUV$ ratio and/or $\tau$ as described above, the only possibility is 
to use the morphological type to obtain a rough estimate of $A(FUV)$.
In Table \ref{afuv_type} we give the average value and standard deviation of $A(FUV)$ obtained 
for our sample in bins of morphological type.
For our sample this method provides an estimate of $A(FUV)$ with an average error of $\sim$0.7 mag.
However we recommend to apply this technique only when accurate an morphological classification is 
available (i.e. the local Universe), otherwise the error on $A(FUV)$ will be considerably larger than the 
one obtained for our sample.

\section{Implications on the study of UV properties of galaxies.}
The simple recipes presented here have the widest scientific application possible, 
requiring only an UV and UV-optical colour.
In order to show their real power, in the following we will 
discuss two different applications of these methods to the study of the UV properties of nearby galaxies and compare them with the age independent techniques usually adopted.

\subsection{Star formation density profiles of resolved galaxies}
\begin{figure*}
\includegraphics[width=8.cm]{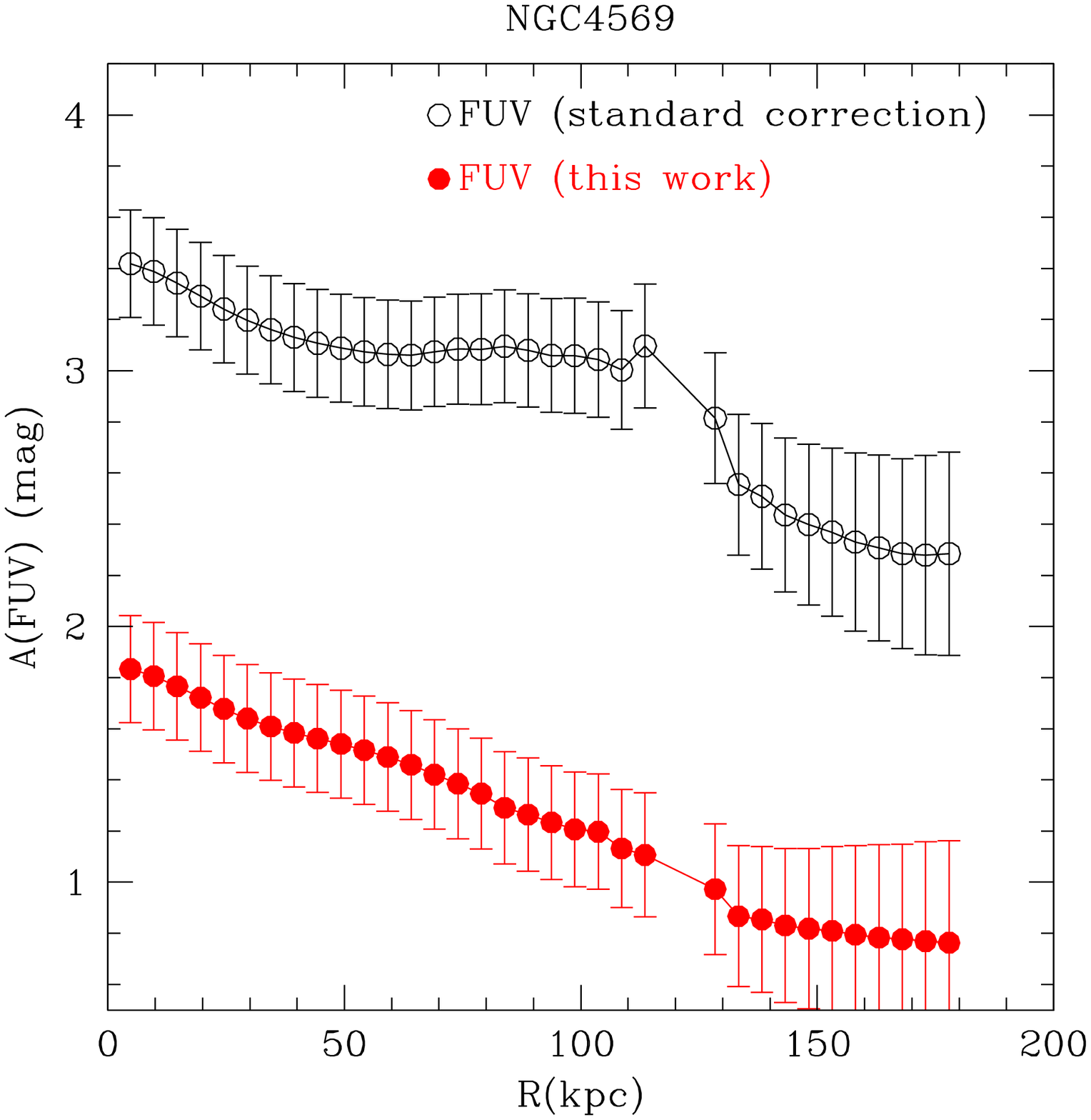}
\includegraphics[width=8.cm]{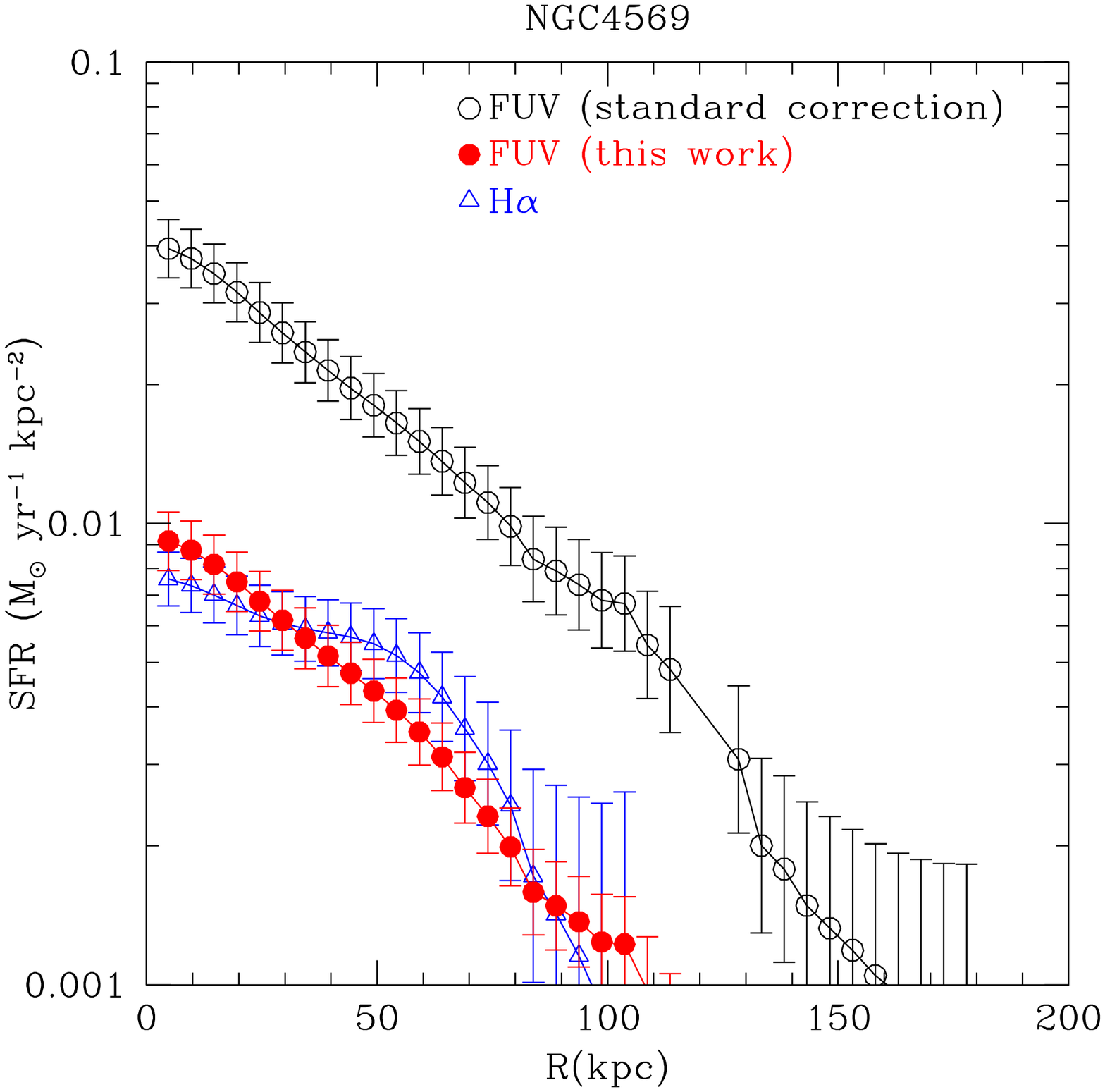}
\caption{\label{n4569} Left: Comparison between the FUV attenuation profile of NGC4569 obtained following the standard $TIR/FUV$ vs. $A(FUV)$ conversion (empty circles) and the one obtained using the conversion presented in this work (filled circles). Right: Star formation rate density profile for NGC4569. Star formation is obtained from the FUV corrected with 
the standard $TIR/FUV$ vs. $A(FUV)$ conversion \citep{buat05} (empty circles), with the recipe presented in this work (filled circles) and from the H$\alpha$ corrected for extinction using the Balmer decrement.}
\end{figure*}

The \emph{age effect} on the $TIR/FUV$ vs. $A(FUV)$ relation is particularly important in panchromatic 
studies of resolved galaxies \citep{samprof04,boissier06,calzetti05,perez06} since in different regions the dust can be heated up by different stellar populations \citep{calzetti05}.\\
We test our method on the nearby spiral galaxy NGC 4569 (M90), the brightest spiral galaxy in
the Virgo cluster recently studied by \cite{n4569}. 
NGC 4569 can be considered as the prototype of HI deficient galaxy  having only about one-tenth of 
the atomic gas of a comparable field galaxy of similar type and size.
This galaxy has a truncated H$\alpha$ and HI  radial profile (at a radius of $\sim$ 5 kpc, \citealp{cayatte94,koop2,n4569}) 
and shows significant colour gradients with star formation activity only in the central 
5 kpc. Therefore, given its significant age gradients, the contribution of UV photons 
to the dust heating probably varies with the distance from the galaxy center, making this object ideal to compare with our recipes with age independent techniques.
As described in \cite{n4569} we determined the $TIR/FUV$ ratio profile combining GALEX-FUV image and Spitzer 24, 70, and 160 $\rm \mu m$ radial profiles.
Then, we use the observed $FUV-H$ colour profile to estimate $\tau$ at each radius, determining 
which $TIR/FUV$ vs. $A(FUV)$ relation to use at each radius.
The $A(FUV)$ profile so obtained is shown in Fig.\ref{n4569} (left panel, empty circles) and compared to the one estimated using the age independent $TIR/FUV$ vs. $A(FUV)$ conversion \citep{buat05}.
As expected, our method gives a FUV attenuation $\sim$1.5 mag lower than the standard conversion implying a factor $\sim$4 difference in the SFR surface density profile obtained from the FUV profile (Fig.\ref{n4569}, right panel).
The results obtained with our technique are supported by the very good agreement 
with the SFR density profile independently obtained from the H$\alpha$ line (corrected 
for extinction using Balmer decrement; Fig. \ref{n4569} right panel, empty triangles). In fact only a recent ($<10^{7}$ yr) starburst $\sim$100 times stronger than the normal star formation rate in NGC~4569 could reconcile 
the difference between the SFR obtained from the H$\alpha$ and from the FUV corrected with the standard methods \citep{jorgehauv}: an extremely unlikely scenario as discussed by 
\cite{vollmer04} and \cite{n4569}.\\
This exercise shows the crucial importance of the age dependent $TIR/FUV$ vs $A(FUV)$ relations 
in the panchromatic study of resolved galaxies.

\subsection{The UV-optical colour magnitude relation of large samples of galaxies.}
\begin{figure*}
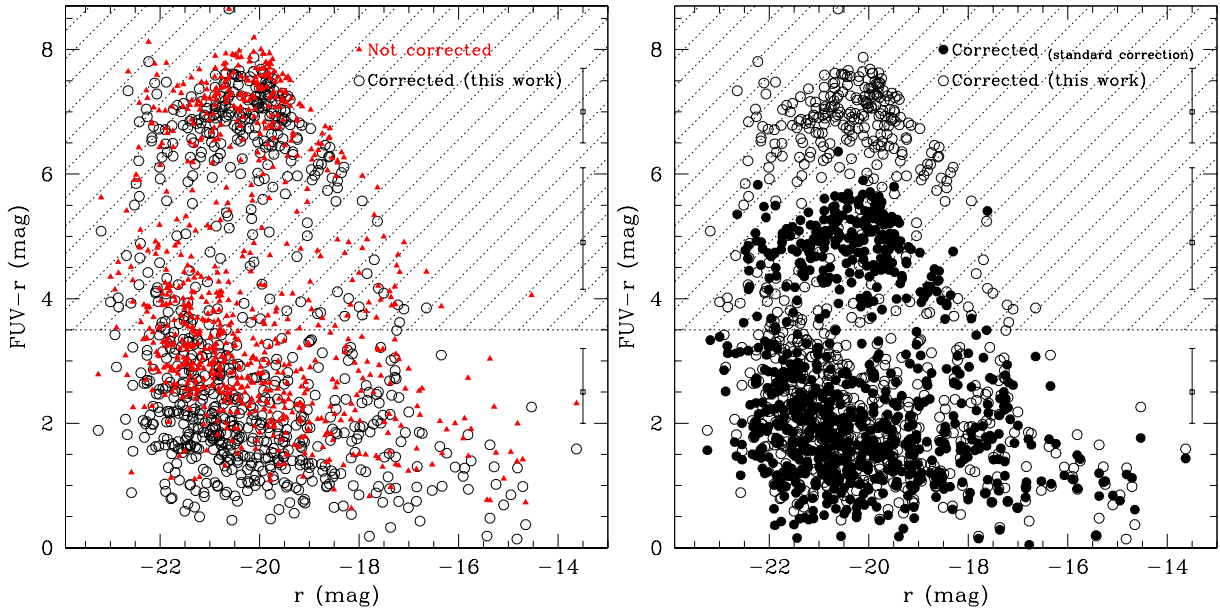

\includegraphics[width=8.cm]{CM2.epsi}
\includegraphics[width=8.cm]{CM1.epsi}
\caption{\label{CM} The $FUV-r$ colour vs. $r$ magnitude relation for galaxies projected on the Coma cluster region.
Left: Triangles indicate the observed (i.e. not corrected for internal extinction) colour-magnitude relation, while empty circles show the colour-magnitude relation after correcting for dust attenuation using the recipes presented in this work. Right: 
Empty circles are as in the right panel whereas filled circles show the colour-magnitude relation obtained using the standard (i.e. age independent) correction. The errors on the estimate of the corrected $FUV-r$ for galaxies in the blue, red sequence and in the transition region are indicated by empty squares.
The unshaded regions indicate the range of observed (i.e. not corrected for internal extinction) colours on which the standard $TIR/FUV$ vs. $A(FUV)$ relation for star-forming galaxies is usually calibrated.}
\end{figure*}

The need of empirical methods to correct for dust attenuation is particular important for the 
study of UV properties of large samples of galaxies lacking far-infrared rest-frame data.
Of particular importance for our understanding of galaxy evolution is the correct interpretation 
of the UV-optical colour magnitude (CM) relation, as recently discussed by \cite{wyder07}, \citep{schiminovich07} and \cite{martin07}.
It is very difficult to use a single correction technique able to deal simultaneously with starbursts, low SSFR objects and elliptical galaxies.
Of particular importance is the correct estimate of $A(FUV)$ for transition objects: galaxies with low SSFR for which the standard corrections calibrated 
on active star forming systems are likely to be not valid. 
In order to quantify the influence of the \emph{age effect} on our interpretation of the 
UV-optical colour magnitude relation we compare our method with the age independent recipes using a sample of galaxies extracted from GALEX observations of the Coma cluster region (GI-Cycle 1, Cortese et al., in preparation). 
This sample includes all galaxies detected in both FUV and NUV bands as well as with SDSS-DR6 
$ugriz$ photometry and with spectroscopic redshift data available (833 galaxies).
Since for this large sample far infrared observations are not available, we use the second method described in Sec.\ref{recipe}, determining the $TIR/FUV$ ratio from the $FUV-NUV$ colour and $\tau$ from the $FUV-i$ colour.
The $FUV-r$ colour-$r$ magnitude relation obtained is compared with the observed one (i.e. not corrected for internal 
extinction) in Fig.\ref{CM} (left panel).  
Whereas galaxies in the blue sequence ($FUV-r\lsim$ 4 mag) have an average FUV dust attenuation $A(FUV)\sim$1.5-2 mag, galaxies 
in the red sequence ($FUV-r\geq$6 mag) show (as expected) a very low amount of attenuation ($A(FUV)\sim$ 0.5 mag) and occupy almost the same parameter space as uncorrected data.
The separation between the red and blue sequences is therefore more evident after the correction 
for dust attenuation.
We remark that our technique should not be blindly applied to elliptical galaxies, since in these objects UV emission 
comes from old stellar populations and not young stars (e.g. \citealp{bosell05}). However if an accurate morphological classification is not 
available the use of the recipes presented here does not introduce a strong systematic bias in the data, as shown in Fig.\ref{CM} (left panel).
This result indicates that our procedures are reliable for old as well as young  stellar populations, contrary 
to previous empirical methods, calibrate and valid only for active star-forming galaxies  (e.g. \citealp{johnson06,salim07}).
This is clearly evident in Fig.\ref{CM} (right panel) where the CM relation obtained with our recipe is compared with the one determined by using 
the age-independent method to convert the $FUV-NUV$ colour into $A(FUV)$ (e.g. \citealp{johnson06,salim07}). 
The difference between the two CM relations is quite impressive: while no significative difference is observed in blue-sequence galaxies, the red sequence shifts $\sim$2 mag towards bluer colours and the gap between the blue and red sequence (i.e. the so-called "green valley", \citealp{wyder07,schiminovich07,martin07}) is considerably reduced. A similar result has been recently shown by \cite{wyder07} when comparing the CM relation determined using the Balmer decrement as indicator of dust 
attenuation with the one estimated using the recipes proposed by \cite{johnson06}.
This result shows how strong the bias can be when blindly using the recipes to estimate $A(FUV)$ without taking into account the \emph{age effect}, leading to an incorrect interpretation of the data and reconstruction of galaxy's evolution history.

\section{Conclusions}
We have investigated the dependence of the $TIR/FUV$ vs. $A(FUV)$ relation on the average age of galaxy stellar populations. 
Our simple method has shown that even for spiral galaxies the use of a standard (i.e. not age dependent) conversion of the $TIR/FUV$ into $A(FUV)$ leads to a systematic 
overestimate (i.e. $\geq$1 mag) of the dust attenuation in galaxies with low specific star-formation, mainly anemic cluster spirals. 
This systematic bias strongly affects our interpretation of UV observations and can produce a significant overestimate (up to a factor 10) 
of the star formation rate, particularly in HI deficient galaxies.
Therefore we have developed different methods for determining the UV dust attenuation taking into account the age dependence of the $TIR/FUV$ vs. $A(FUV)$ relation.
These recipes require only an UV colour and an UV-optical/near infrared colour providing an estimate 
of the UV attenuation with an average uncertainty varying from 
$\sim$+0.1/-0.25 mag for $\tau\geq$6 Gyr ($FUV-H<$5 mag) to a maximum of +0.4/-0.8 mag for $3.5\leq\tau\leq4.5$ Gyr (7$<FUV-H<$8.5 mag), when far-infrared observations are available, and from $\sim$+0.5/-0.7 mag for $\tau\geq$6 Gyr ($FUV-H<$5 mag) to a maximum of $\sim$+1/-1.2 mag for $3.5\leq\tau\leq4.5$ Gyr (7$<FUV-H<$8.5 mag) when far-infrared data are not available.
The small amount of multiwavelength data necessary for their application makes these methods extremely useful for the widest possible range of application eventually providing us with a less biased view of the UV properties of galaxies in the local Universe and at higher redshift. 

\section*{Acknowledgments}
We want to thank the anonymous referee, whose comments and suggestions were extremely useful for improving the present manuscript. 
We thank Jonathan Davies for useful discussions and comments on this manuscript. 
LC is supported by the UK Particle Physics and Astronomy Research Council. 
GALEX (Galaxy Evolution Explorer) is a NASA Small Explorer, launched in April 2003.
We gratefully acknowledge NASA's support for construction, operation,
and science analysis for the GALEX mission,
developed in cooperation with the Centre National d'Etudes Spatiales
of France and the Korean Ministry of Science and Technology.
This research has made use of the NASA/IPAC Extragalactic Database, which is operated by the Jet Propulsion Laboratory, California Institute of Technology, under contract to NASA and of the GOLDMine database

\section*{Appendix A: The dependence of the $TIR/FUV$ vs $A(FUV)$ relation on extinction law, 
dust geometry and star formation history. }
Are the results obtained in this work and the recipes proposed to estimate the UV dust attenuation 
model-dependent? In order to answer to this question, in this section we investigate the dependence of our results on the three free parameters entering our model: the shape of the extinction law, the dust geometry and the star formation history.

\begin{figure*}
\includegraphics[width=8.cm]{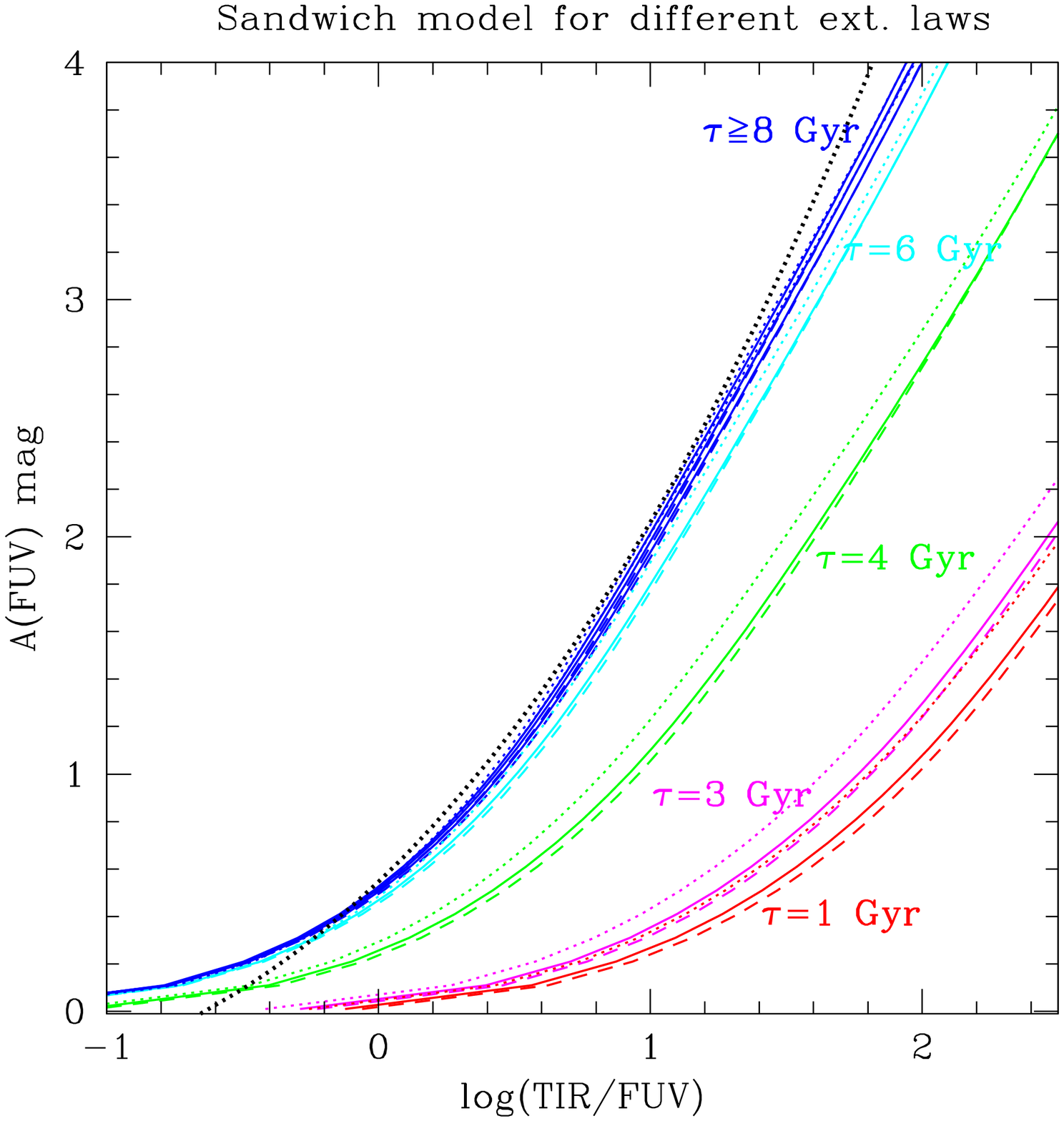}
\includegraphics[width=8.cm]{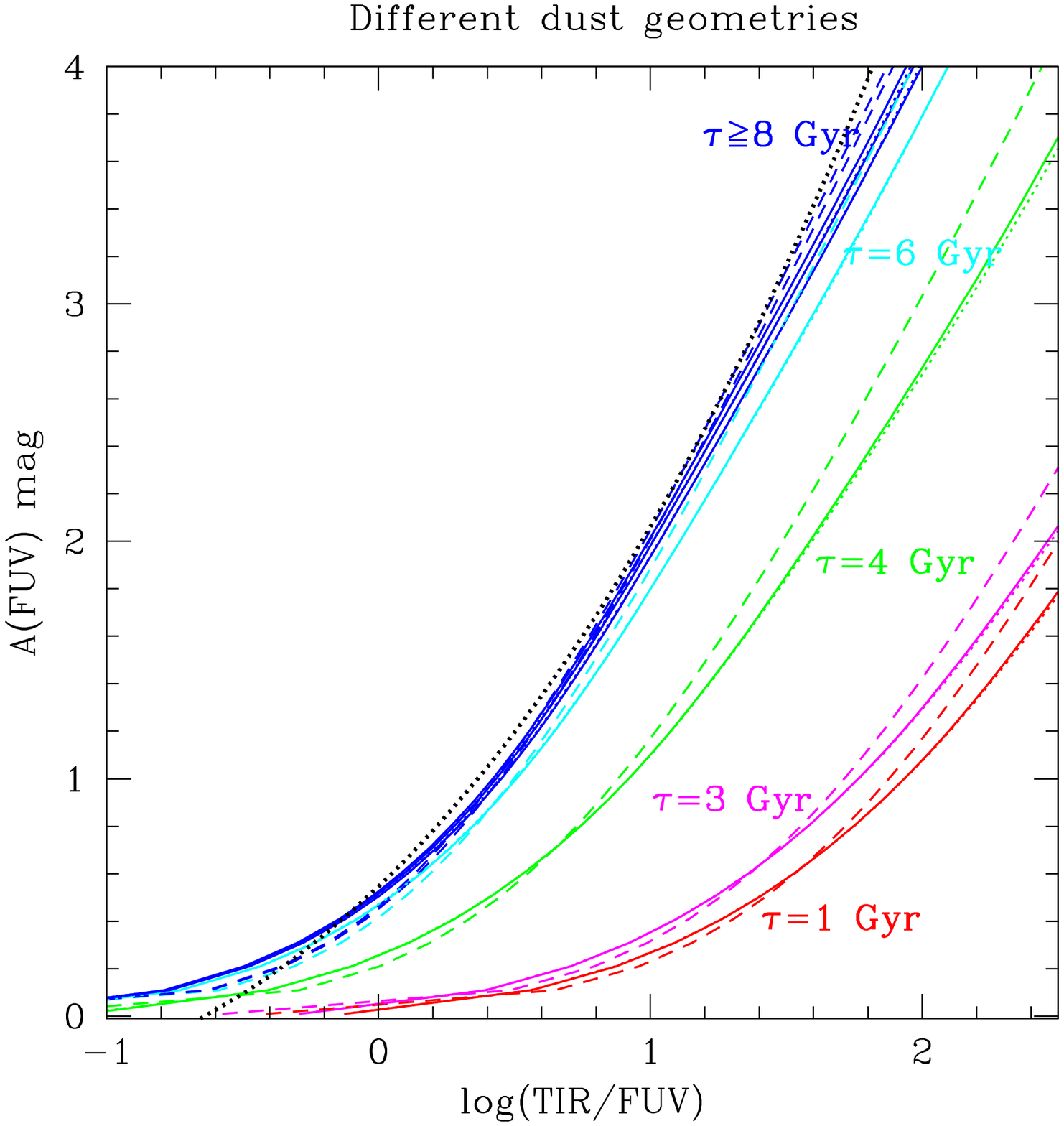}
\caption{\label{modgeo} The relationship between the $TIR/FUV$ ratio and the 
FUV attenuation $A(FUV)$ obtained from our model using different 
extinction laws and geometries. Left Panel:  Sandwich geometry with 
LMC (solid line), Milky Way (dashed) and Small Magellanic cloud (dotted line) extinction laws. Right: LMC extinction law with Sandwich (solid line), simple Slab (dotted) geometries and 
Calzetti attenuation law (dashed line).
In both panels the dotted black line indicates the age independent relation given by Buat et al. (2005).}
\end{figure*}

In Fig.\ref{modgeo} (Left) we show the $TIR/FUV$ vs $A(FUV)$ relation obtained for different values $\tau$, a Sandwich geometry and three different extinction laws: LMC (solid line), Milky Way (dotted line) and Small Magellanic Cloud (dashed line).
As already pointed out by several authors (e.g. \citealp{witt00,buat05}), the $TIR/FUV$ vs $A(FUV)$ relation is almost independent of the extinction law with a typical variation limited to  $\leq$0.2 mag, considerably lower than the effect of $\tau$ on the 
 $TIR/FUV$ vs $A(FUV)$ relation.
 
 Similar results are obtained if we investigate the effect of different dust geometries.
 In Fig.\ref{modgeo} (Right) we compare the results obtained for the sandwich model (solid line) with 
 a simple slab geometry (\citealp{disney89}, dashed line). We also added the results for a 
 \cite{calzetti94} attenuation law (dotted line).   
Even in this case, we observed no dependence of the $TIR/FUV$ vs. $A(FUV)$ relation on the dust geometry within $\leq$0.2 mag.
Similar results are obtained if we use the homogeneous and clumpy dust models proposed by \cite{witt00} (not shown).

\begin{figure}
\centering
\includegraphics[width=8.48cm]{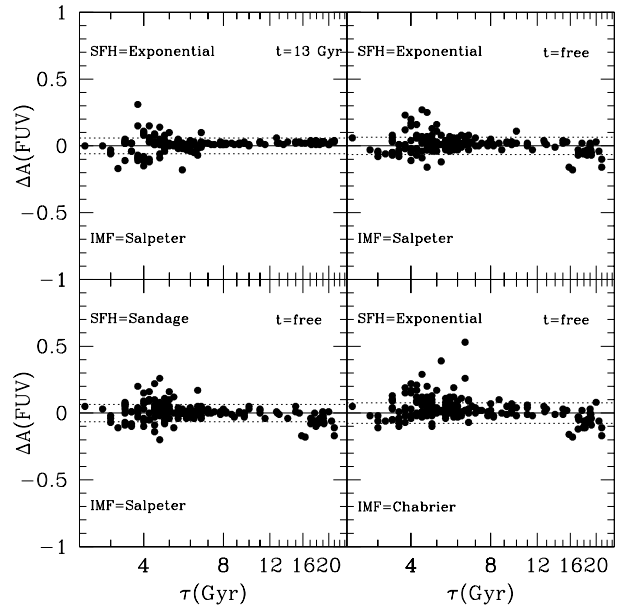}
\caption{\label{modsfh} Difference in the estimate of $A(FUV)$ with the SED technique described 
in Sec.\ref{data} ($t=$13 Gyr, 'a la Sandage' SFH and Salpeter IMF) and various combination of SFH (Sandage or exponential), age (variable or fixed to 13 Gyr) and initial mass function (IMF, Salpeter or Chabrier). The dashed lines indicates the 1$\sigma$ difference in the estimate of $A(FUV)$. }
\end{figure}

Finally, we fitted the 191 galaxies in our sample with various SED libraries in order to test the dependence of the results presented in Sec.\ref{data} and  Sec.\ref{recipe} from SFH and age. 
The SED library used in Sec.\ref{model} has been obtained by fixing the shape of the 
SFH (assumed to be 'a la Sandage'), the age of the galaxy (assumed equal to 13 Gyr) and the 
stellar initial mass function (IMF, assumed to be a Salpeter IMF).
We produced different set of SED libraries varying every time one of these three free parameters and we fitted 
them to the data. In particular we considered an exponential SFH, a fixed ($t$=13 Gyr) or free (in the range 0$<t<$15 Gyr) age and 
a \cite{chabrier} IMF. In Fig.\ref{modsfh} is presented the difference in the estimate of $A(FUV)$ between the library adopted in Sec.\ref{model} and some of the various combination adopted.
For each combination the standard deviation in the estimate of $A(FUV)$ is lower than 0.1-0.2 mag.
This result was somehow expected since the $TIR/FUV$ vs. $A(FUV)$ relation is only affected by 
the actual shape of the SED and not by the shape of the past (older than 1-2 Gyr) SFH.
We remark that our results are not applicable only in the case of an embedded starburst in 
a galaxy with a second older, less embedded stellar population \citep{gordon00}. 
At UV and IR wavelengths, the starburst would dominate, but at optical and near-IR wavelengths the older population would dominate.
However, at least for spiral galaxies in the local Universe this seems not to be the case \citep{jorgehauv} and smooth SFH like 'a la Sandage' 
and exponential well reproduce the observations (e.g. \citealp{boselli,gav02,heavens,panter07})

We can therefore conclude that the results obtained from our model and the recipes presented in this 
work are model-independent at least within $\sim$0.2 mag. This uncertainty is still lower or equal to 
the uncertainty in the estimate of $A(FUV)$ from our recipes.
%The lowest observational error in the estimate of the $TIR/FUV$ ratio is in fact $\sim$0.2 dex, implying an average uncertainty of $\sim$ 0.3 mag in the estimate of $A(FUV)$ (see Fig.\ref{obserr}).
%A comparison between Figs.\ref{modgeo} and Fig.\ref{obserr} clearly shows that in general the observational errors dominate the uncertainty due to the assumptions on which our model is based.

\section*{Appendix B: Recipes to determine $A(NUV)$ when FUV observations are not available.}
In this Appendix we discuss the possibility to extend our recipes to the NUV band in the case that GALEX-FUV 
observations are available. The GALEX-NUV filter ($\rm \lambda_{eff}=2310\AA, \Delta \lambda=1000\AA$) lies in a region 
where the SED of galaxies with $\tau <$ 4 Gyr is strongly dependent on stellar metallicity for $Z>0.2~ \rm Z_{\odot}$ (e.g. see Fig. 6 in \citealp{gav02}). This strongly affects the $TIR/NUV$ vs $A(NUV)$ relation which, for low values of $\tau$, can vary of even 0.4 mag from 
$Z=0.2~ \rm Z_{\odot}$ to $Z=0.02~ \rm Z_{\odot}$, apparently complicating the use of our model directly on the NUV band. However for none of the galaxies with $\tau <$ 4 Gyr in our sample the best fitting model has metallicity lower than $Z=0.2~ \rm Z_{\odot}$, consistent with the fact that in the local Universe evolved galaxies tend to be metal rich (e.g. \citealp{gallazzi05}).  Therefore for $\tau <$ 4 Gyr we compute the average $TIR/NUV$ vs $A(NUV)$ relation by combining only the relations obtained in the metallicity range $0.02<Z<2.5~ \rm Z_{\odot}$. This makes our 
recipes not valid for $\tau <$ 4 Gyr and $Z<0.2~ \rm Z_{\odot}$.
The average $TIR/NUV$ vs. $A(NUV)$ relations for different values of $\tau$ so obtained are presented in Table \ref{convNUV}.
In Table \ref{tauNUV} are given the relations to use to estimate $\tau$ from the NUV-optical/near-infrared colours. Their typical dispersion is 
consistent with the one observed in the FUV-optical/near-infrared colours vs. $\tau$ relations.
As discussed in Sec.\ref{recipe} the suggested recipes to determine $A(NUV)$ result:\\

\noindent
i) {\bf Far-infrared observations are available.} Determine the observed $TIR/NUV$ ratio and use one of the relations  presented in Table \ref{tauNUV} to determine $\tau$. Finally use the value of $\tau$ obtained with this procedure to choose the conversion between the $TIR/FUV$ ratio and $A(FUV)$ from the relations provided in Table \ref{convNUV}. The typical error on the estimate of $A(NUV)$ (computed through a Montecarlo simulation similar to the one described in Sec. 4) varies between $\sim$+0.1/-0.2 mag for $\tau\geq$8 Gyr, to +0.3/-0.7 mag for $3.5\leq\tau\leq4.5$ Gyr.\\

\noindent
ii) {\bf Far-infrared observations are not available.} In this case the first step is 
to find a way to determine the $TIR/NUV$ from the observations available. 
Once the $TIR/NUV$ ratio has been determined it is sufficient to proceed as described in the previous point: i.e. estimate $\tau$ from an ultraviolet-optical/near infrared colour and than convert $TIR/NUV$ in $A(NUV)$ using the proper 
relation in Table \ref{convNUV}. 
The typical error on the estimate of $A(NUV)$ (computed through a Montecarlo simulation similar to the one described in Sec. 4) varies between $\sim$+0.4/-0.6 mag for $\tau\geq$8 Gyr, and to $\sim$+0.9/-1 mag for $3.5\leq\tau\leq4.5$ Gyr.\\

\noindent
In Table \ref{anuv_type} are presented the average value of $A(NUV)$ obtained for the different morphological types in our sample.
The morphological type can be used to estimate $A(NUV)$ only if the to previous methods can not be applied and an accurate morphological classification is available.

\onecolumn

%\begin{figure}
%\centering
%\includegraphics[width=8.48cm]{AFUV_obserr.ps}
%\caption{\label{obserr} The $TIR/FUV$ vs. $A(FUV)$ relation as shown in Fig.\ref{tauafuv}. The dashed regions indicate the uncertainty in the estimate of $A(FUV)$ given an observational error of $\sim$0.2 dex in the estimate of the $TIR/FUV$ ratio.  }
%\end{figure} 

\clearpage

\newpage

\begin{table}
\caption {Relations to convert the $TIR/FUV$ ratio in $A(FUV)$ for different values of $\tau$ and FUV-near-infrared/optical colours.}
\[
\label{convFUV}
\begin{array}{ccccccccccc}
\hline
\noalign{\smallskip}
\multicolumn{11}{c}{A(FUV)=a1+a2\times x+a3\times x^{2}+a4 \times x^{3}+ a5 \times x^{4} ~~where~ x=\log(TIR/FUV)}\\
\noalign{\smallskip}
\hline
\tau \rm (Gyr) &     a1         &      a2      &      a3       &      a4      &      a5   & FUV-H & FUV-i & FUV-r & FUV-g & FUV-B    \\  
\noalign{\smallskip}									       
\hline											       
\noalign{\smallskip}									     	  
\leq2.6	  &	0.02025 & 0.06107 & 0.07212 & 0.10588 & -0.01517 &	  10.5  &   7.5  &    7.3  &	6.7  &    6.3  \\						   
2.8	  &	0.02355 & 0.06934 & 0.08725 & 0.10339 & -0.01526 &	  10.0  &   7.0  &    6.9  &	6.3  &	  5.9  \\						   
3	  &	0.03404 & 0.09645 & 0.12452 & 0.09679 & -0.01548 &	  9.6   &   6.6  &    6.5  &	5.9  &	  5.6  \\						   
3.2	  &	0.05822 & 0.15524 & 0.17801 & 0.08664 & -0.01593 &	  9.2   &   6.2  &    6.1  &	5.6  &	  5.2  \\						   
3.4	  &	0.09944 & 0.24160 & 0.23161 & 0.07580 & -0.01671 &	  8.8   &   5.9  &    5.8  &	5.3  &	  4.9  \\						   
3.6	  &	0.15293 & 0.33799 & 0.27713 & 0.06638 & -0.01792 &	  8.4   &   5.5  &    5.4  &	5.0  &	  4.6  \\						   
3.8	  &	0.20982 & 0.42980 & 0.31431 & 0.05909 & -0.01957 &	  8.1   &   5.2  &    5.1  &	4.7  &	  4.3  \\						   
4	  &	0.26302 & 0.51013 & 0.34522 & 0.05377 & -0.02164 &	  7.8   &   4.9  &    4.8  &	4.4  &	  4.1  \\						   
4.2	  &	0.30899 & 0.57732 & 0.37157 & 0.05000 & -0.02399 &	  7.5   &   4.6  &    4.6  &	4.2  &	  3.8  \\						   
4.4	  &	0.34695 & 0.63224 & 0.39438 & 0.04739 & -0.02650 &	  7.2   &   4.3  &    4.3  &	3.9  &	  3.6  \\						   
4.6	  &	0.37760 & 0.67674 & 0.41420 & 0.04555 & -0.02900 &	  6.9   &   4.1  &    4.0  &	3.7  &	  3.3  \\						   
4.8	  &	0.40210 & 0.71272 & 0.43139 & 0.04426 & -0.03140 &	  6.6   &   3.8  &    3.8  &	3.5  &	  3.1  \\						   
5	  &	0.42168 & 0.74191 & 0.44624 & 0.04332 & -0.03362 &	  6.3   &   3.6  &    3.6  &	3.3  &	  2.9  \\						   
5.4	  &	0.45013 & 0.78536 & 0.47009 & 0.04210 & -0.03745 &	  5.8   &   3.1  &    3.1  &	2.9  &	  2.5  \\						   
5.8	  &	0.46909 & 0.81520 & 0.48787 & 0.04138 & -0.04050 &	  5.4   &   2.7  &    2.7  &	2.5  &	  2.1  \\						   
6.2	  &	0.48223 & 0.83642 & 0.50127 & 0.04092 & -0.04288 &	  5.0   &   2.3  &    2.3  &	2.1  &	  1.8  \\						   
6.6	  &	0.49167 & 0.85201 & 0.51152 & 0.04060 & -0.04475 &	  4.6   &   1.9  &    2.0  &	1.8  &	  1.5  \\						   
   7     &	0.49867 & 0.86377 & 0.51952 & 0.04038 & -0.04624 &	  4.2   &   1.6  &    1.6  &	1.5  &	  1.1  \\						   
%7.4	  &	0.50401 & 0.87287 & 0.52586 & 0.04022 & -0.04744 &	  3.8   &   1.2  &    1.3  &	1.2  &	  0.9  \\						   
\geq8 &	0.50994 & 0.88311 & 0.53315 & 0.04004 & -0.04883 &     <4.0 &   <1.2 &  <1.3  &   <1.2  &	<1.0  \\						   
\noalign{\smallskip}
\hline
\end{array}
\]
Each value of $\tau$ has been converted into FUV-near-infrared/optical colours using the relations presented in Table \ref{taucol}
\end{table}

\begin{table}
\caption {Linear relations useful to estimate $\tau$ from observed (i.e. not corrected for dust attenuation) far ultraviolet-near-infrared/optical colours ($\log(\tau)=a\times x+b$). These relations 
have been calibrated in the range $2\leq\tau\leq7$ Gyr.
}
\[
\label{taucol}
\begin{array}{ccccc}
\hline
\noalign{\smallskip}
x            & a & b & r  & validity~range\\
\noalign{\smallskip}
\hline
\noalign{\smallskip}									       
FUV-H  & -0.068 & 1.13 & -0.91  & 4.5\lsim FUV-H\\    
FUV-i    & -0.073  & 0.96 & -0.91 & 1.6\lsim FUV-i\\
FUV-r    & -0.076 & 0.97 & -0.91  & 1.6\lsim FUV-r\\
FUV-g   & -0.083  & 0.97 & -0.91 & 1.5\lsim FUV-g\\
FUV-B  & -0.083  & 0.94 & -0.91 &  1.1\lsim FUV-B\\

\noalign{\smallskip}
\hline
\end{array}
\]
\end{table}

\begin{table}
\caption {The average value of $A(FUV)$ for our sample in bins of morphological type.}
\[
\label{afuv_type}
\begin{array}{ccc}
\hline
\noalign{\smallskip}
type            & A(FUV) & \sigma  \\
\noalign{\smallskip}
\hline
\noalign{\smallskip}									       
Sa    &  1.5 & 0.9 \\
Sab  &  1.6 & 0.6 \\
Sb    &  1.8 & 0.7 \\
Sbc  &  1.8 & 0.6 \\
Sc    &  1.5 & 0.6 \\
Scd  &  1.2 & 0.8 \\
Sd    &  0.9 & 0.6 \\
Sm   &  1.5 & 0.4 \\
Im-BCD & 1.7 & 0.7 \\
\noalign{\smallskip}
\hline
\end{array}
\]
\end{table}

\begin{table}
\caption {Relations to convert the $TIR/NUV$ ratio in $A(NUV)$ for different values of $\tau$ and NUV-near-infrared/optical colours.}
\[
\label{convNUV}
\begin{array}{ccccccccccc}
\hline
\noalign{\smallskip}
\multicolumn{11}{c}{A(NUV)=a1+a2\times x+a3\times x^{2}+a4 \times x^{3}+ a5 \times x^{4} ~~where~ x=\log(TIR/NUV)}\\
\noalign{\smallskip}
\hline
\tau \rm (Gyr) &     a1         &      a2      &      a3       &      a4      &      a5   & NUV-H & NUV-i & NUV-r & NUV-g & NUV-B    \\  
\noalign{\smallskip}									       
\hline											       
\noalign{\smallskip}									     	  
\leq2.6	   &	0.04030 & 0.12924 & 0.02301 & 0.14889  & -0.01909  & 	    9.3 &      6.2 & 	  5.9 &      5.3 & 	5.0	\\		
2.8	   &	0.04436 & 0.13118 & 0.03649 & 0.14669  & -0.01911  & 	    8.9 &      5.9 & 	  5.6 &      5.0 & 	4.7	\\		
3	   &	0.05370 & 0.13840 & 0.06591 & 0.14195  & -0.01921  & 	    8.5 &      5.5 & 	  5.3 &      4.7 & 	4.4	\\		
3.2	   &	0.07037 & 0.15838 & 0.11195 & 0.13508  & -0.01962  & 	    8.2 &      5.2 & 	  5.0 &      4.5 & 	4.2	\\		
3.4	   &	0.09423 & 0.19557 & 0.16580 & 0.12827  & -0.02060  & 	    7.9 &      4.9 & 	  4.7 &      4.2 & 	3.9	\\		
3.6	   &	0.12309 & 0.24566 & 0.21875 & 0.12335  & -0.02236  & 	    7.5 &      4.6 & 	  4.5 &      4.0 & 	3.7	\\		
3.8	   &	0.15343 & 0.30012 & 0.26696 & 0.12089  & -0.02490  & 	    7.3 &      4.4 & 	  4.2 &      3.8 & 	3.5	\\		
4	   &	0.18195 & 0.35210 & 0.30967 & 0.12051  & -0.02805  & 	    7.0 &      4.1 & 	  4.0 &      3.6 & 	3.3	\\		
4.2	   &	0.21793 & 0.41714 & 0.35825 & 0.11985  & -0.03139  & 	    6.7 &      3.9 & 	  3.8 &      3.4 & 	3.1	\\		
4.4	   &	0.23640 & 0.45238 & 0.38817 & 0.12220  & -0.03488  & 	    6.5 &      3.7 & 	  3.5 &      3.2 & 	2.9	\\		
4.6	   &	0.25119 & 0.48147 & 0.41412 & 0.12486  & -0.03822  & 	    6.2 &      3.5 & 	  3.3 &      3.0 & 	2.7	\\		
4.8	   &	0.26292 & 0.50518 & 0.43641 & 0.12761  & -0.04131  & 	    6.0 &      3.2 & 	  3.1 &      2.9 & 	2.5	\\		
5	   &	0.27220 & 0.52449 & 0.45540 & 0.13025  & -0.04406  & 	    5.8 &      3.0 & 	  3.0 &      2.7 & 	2.3	\\		
5.4	   &	0.28548 & 0.55319 & 0.48523 & 0.13492  & -0.04854  & 	    5.3 &      2.7 & 	  2.6 &      2.4 & 	2.0	\\		
5.8	   &	0.29415 & 0.57274 & 0.50686 & 0.13873  & -0.05189  & 	    5.0 &      2.3 & 	  2.3 &      2.1 & 	1.7	\\		
6.2	   &	0.30003 & 0.58647 & 0.52278 & 0.14176  & -0.05439  & 	    4.6 &      2.0 & 	  2.0 &      1.8 & 	1.5	\\		
6.6	   &	0.30418 & 0.59643 & 0.53473 & 0.14418  & -0.05626  & 	    4.3 &      1.7 & 	  1.7 &      1.6 & 	1.2	\\		
        7  &	0.30720 & 0.60385 & 0.54390 & 0.14613  & -0.05770  & 	    3.9 &      1.4 & 	  1.4 &      1.3 & 	1.0	\\		
%7.4	   &	0.30947 & 0.60955 & 0.55107 & 0.14769  & -0.05882  & 	    3.6 &      1.1 & 	  1.2 &      1.1 & 	0.8	\\		
\geq8  &	0.31194 & 0.61587 & 0.55922 & 0.14955  & -0.06009  & 	  <3.6 &    <1.1 & 	<1.2 &   <1.1 & <0.8	\\		
\noalign{\smallskip}
\hline
\end{array}
\]
Each value of $\tau$ has been converted into NUV-near-infrared/optical colours using the relations presented in Table \ref{tauNUV}.
\end{table}

\begin{table}
\caption {Linear relations useful to estimate $\tau$ from observed (i.e. not corrected for dust attenuation) near ultraviolet-near-infrared/optical colours ($\log(\tau)=a\times x+b$). These relations 
have been calibrated and are valid only in the range $2\leq\tau\leq7$ Gyr}
\[
\label{tauNUV}
\begin{array}{ccccc}
\hline
\noalign{\smallskip}
x            & a & b & r  & validity~range\\
\noalign{\smallskip}
\hline
\noalign{\smallskip}									       
NUV-H  & -0.080  & 1.16 & -0.88 & 4.0\lsim NUV-H\\    
NUV-i   & -0.089   & 0.97 & -0.88 & 1.5\lsim NUV-i\\
NUV-r   & -0.095   & 0.98 & -0.88 & 1.5\lsim NUV-r\\
NUV-g & -0.108    & 0.99 & -0.89 & 1.5\lsim NUV-g\\
NUV-B & -0.107   & 0.95  & -0.90 & 1.0\lsim NUV-B\\
\noalign{\smallskip}
\hline
\end{array}
\]
\end{table}

\begin{table}
\caption {The average value of $A(NUV)$ for our sample in bins of morphological type.}
\[
\label{anuv_type}
\begin{array}{ccc}
\hline
\noalign{\smallskip}
type            & A(NUV) & \sigma  \\
\noalign{\smallskip}
\hline
\noalign{\smallskip}									       
Sa    &  1.2 & 0.7 \\
Sab  &  1.3 & 0.5 \\
Sb    &  1.4 & 0.6 \\
Sbc  &  1.4 & 0.5 \\
Sc    &  1.2 & 0.5 \\
Scd  &  0.9 & 0.7 \\
Sd    &  0.8 & 0.5 \\
Sm   &  1.2 & 0.3 \\
Im-BCD & 1.3 & 0.6 \\
\noalign{\smallskip}
\hline
\end{array}
\]
\end{table}

\end{document}